\documentclass[11pt,a4paper]{article}

\usepackage[T1]{fontenc}
\usepackage[utf8]{inputenc}
\usepackage{lmodern}
\usepackage[margin=2.5cm]{geometry}
\usepackage{hyperref}
\usepackage{booktabs}
\usepackage{tabularx}
\usepackage{longtable}
\usepackage{array}
\usepackage{listings}
\usepackage{xcolor}
\usepackage{enumitem}
\usepackage{microtype}
\usepackage{titlesec}
\usepackage{abstract}
\usepackage{parskip}
\usepackage{amsmath}
\usepackage{caption}
\usepackage{float}
\usepackage{textcomp}
\usepackage{bookmark}

\definecolor{codebg}{RGB}{248,248,248}
\definecolor{codeframe}{RGB}{200,200,200}
\definecolor{codecomment}{RGB}{100,100,100}
\definecolor{codekw}{RGB}{0,100,180}
\definecolor{codestring}{RGB}{160,32,32}
\definecolor{gray}{RGB}{160,160,160}

\lstdefinelanguage{Tcl}{
  morekeywords={proc,set,return,if,else,elseif,foreach,for,while,
                namespace,eval,variable,upvar,uplevel,lindex,lset,
                lappend,lrange,llength,list,dict,string,expr,incr,
                catch,try,throw,trap,finally,error,puts,load,source,
                package,require,voo,class,method,constructor,
                importMethods,public,private,oo,itcl},
  sensitive=true,
  morecomment=[l]{\#},
  morestring=[b]",
  morestring=[b]{},
}

\hypersetup{
  colorlinks=true,
  linkcolor=blue!60!black,
  citecolor=green!50!black,
  urlcolor=blue!70!black,
  pdftitle={Vanilla Object Orientation (VOO): A Value-Semantics Approach to Classes in Tcl},
  pdfauthor={Alan Araujo},
}

\title{\textbf{Vanilla Object Orientation (VOO)}: \\
       A Value-Semantics Approach to Classes in Tcl}
\author{Alan Araujo \\
        \small \href{mailto:aarajo@cadence.com}{aarajo@cadence.com}\ |\ %
               \href{mailto:aaraujo.articles@gmail.com}{aaraujo.articles@gmail.com} \\
        \small Cadence Design Systems, Inc.}
\date{}

\newcolumntype{L}[1]{>{\raggedright\arraybackslash}p{#1}}

\begin{document}

\maketitle
\thispagestyle{empty}

\begin{abstract}
I present \textbf{Vanilla Object Orientation (VOO)}, a framework that composes classes from
Tcl's native data structures---lists and dictionaries---rather than introducing additional
framework infrastructure. VOO objects are plain Tcl lists with automatic memory management
through copy-on-write semantics, eliminating the destructor burden inherent in TclOO and Itcl.
Benchmarks on Tcl~8.6.13 and Tcl~9.0 show VOO achieves 7--18$\times$ faster object creation
and 4--6$\times$ superior memory efficiency compared to TclOO\@. A companion C++ migration
path (\textbf{VOO~C++}) further improves field-access speed (setter 2.3--2.6$\times$ faster)
and memory (6.8--9.8$\times$ lighter than TclOO), while preserving an identical Tcl call-site
API\@. Cross-version analysis confirms that VOO's compositional design scales better than
framework-based approaches as the interpreter evolves.
\end{abstract}


\section{Introduction}

The integration of object orientation into Tcl has spanned more than two decades,
characterized by ongoing tension between expressive power and philosophical consistency.
\textbf{Incr Tcl (Itcl)}~\cite{dejong,tclwiki-itcl}, developed in the 1990s, introduced
traditional OO features but created fundamental semantic conflicts with the core language.
The community debated standardization through several Tcl Improvement Proposals (TIPs)~%
\cite{tclwiki-tip257}: TIP~\#6 (rejected---include Itcl in core), TIP~\#50 (bundle Itcl
without integration), and TIP~\#257 (TclOO in Tcl~8.6). TIP~\#257~\cite{fellows2006},
led by Donal Fellows, implemented \textbf{TclOO} as a purpose-built OO system for core
integration. Released with Tcl~8.6 in 2013~\cite{tcl86}, TclOO addressed many
compatibility challenges while retaining reference semantics that require explicit object
destruction.

\textbf{Vanilla Object Orientation (VOO)} represents the next stage in this evolution.
Rather than importing OO paradigms from other languages, VOO composes classes from Tcl's
established patterns---lists, dictionaries, namespaces, and procedures. The fundamental
question motivating this work is: \emph{How can object orientation be integrated without
compromising the characteristics that define Tcl's identity?} This question manifests
through several tensions: reference vs.\ value semantics, framework infrastructure
vs.\ compositional patterns, feature completeness vs.\ philosophical alignment, and syntax
familiarity vs.\ conceptual simplicity. The optimal OO framework for Tcl is one that
appears as a natural extension---amplifying the language's inherent strengths rather than
supplanting them.

This paper makes six contributions:
\begin{enumerate}[leftmargin=*]
  \item A novel architecture composing classes from native data structures with automatic
        memory management.
  \item Type-aware field declarations with zero runtime overhead.
  \item Comprehensive benchmarking on Tcl~8.6.13 and~9.0 demonstrating 7--18$\times$
        faster creation and 4--6$\times$ better memory than TclOO, extended to VOO~C++.
  \item The first cross-version scalability analysis of Tcl OO frameworks.
  \item A seamless migration path from lists $\to$ VOO objects $\to$ C++.
  \item Design principles showing that simplicity and performance are complementary
        objectives.
\end{enumerate}

\section{Background and Related Work}

\subsection{Tcl's Core Data Model}

Tcl's \texttt{Tcl\_Obj} structure maintains both a string representation and an internal
representation, with reference counting for automatic memory management and copy-on-write
(COW) for shared references. Lists and dictionaries are highly optimized built-in types
that exploit this infrastructure. VOO leverages these existing mechanisms rather than
reimplementing them.

\subsection{Itcl and TclOO}

Itcl introduced OO to Tcl as a separate system layered atop the language, resulting in
fundamental incompatibilities: its \texttt{variable} command conflicted with Tcl's
namespace \texttt{variable}; its access control was bypassable through namespace
manipulation; class redefinition was blocked, violating Tcl's dynamic nature; and
multiple incompatible destruction mechanisms existed. These challenges originated from
designing OO constructs independently of Tcl's philosophical principles.

TclOO (TIP~\#257)~\cite{fellows2006} was a deliberate departure from Itcl, with a new
\texttt{oo::} namespace, a minimalist core, improved namespace alignment, and unified
destruction via \texttt{destroy}. However, persistent limitations remain: objects are
handles requiring explicit destruction, omitted \texttt{destroy} calls cause memory
leaks, each object maintains a procedure handle with computational and memory overhead,
and converting lists/dicts to TclOO objects requires API changes.

\subsection{Lessons from Two Decades}

The evolution reveals that philosophical consistency is critical, clean architectural
breaks can be justified, extensibility enables ecosystem growth, and migration paths
are essential. VOO applies these lessons by maximizing philosophical alignment,
providing incremental migration paths, and prioritizing simplicity over comprehensive
features.

\section{Design Philosophy}

VOO's fundamental design decision is that \textbf{objects are values, not references}.
A VOO object is a plain Tcl list---creating a copy with \texttt{set p2 \$p1} shares the
reference via copy-on-write, and modifying \texttt{p2} leaves \texttt{p1} unchanged.
This eliminates destructor management entirely: objects are garbage-collected when their
reference count reaches zero, preventing the memory leak vulnerabilities inherent in
manual destruction.

Classes are implemented as standard Tcl namespaces. Field indices are stored as
namespace variables, and accessors are ordinary procedures. No specialized infrastructure
is required. VOO constructs OO capabilities from established patterns: lists for object
storage, dictionaries for map fields, namespaces for encapsulation, procedures for
methods, \texttt{upvar} for setter-by-reference, and Tcl's COW for automatic memory
management. This yields seamless integration with existing codebases, predictable
performance from well-understood primitives, and simplified debugging through direct
string representation.

Performance advantages follow directly from this simplicity. Getters are a single
\texttt{lindex} with a precomputed index; setters use \texttt{upvar} and \texttt{lset}
for in-place modification; objects carry no wrapper overhead. Copy-on-write is delegated
entirely to Tcl's existing mechanisms.

\begin{lstlisting}[language=Tcl, caption={VOO getter and setter --- the entire runtime implementation}]
# VOO getter and setter --- the entire runtime implementation
proc Point::get.x {this} { return [lindex $this 0] }
proc Point::set.x {thisVar value} { upvar $thisVar this; lset this 0 $value }
\end{lstlisting}

\section{VOO Framework Design}

\subsection{Type-Aware Fields and Constructors}

VOO provides expressive field declarations with type annotations (\texttt{double\_t},
\texttt{int\_t}, \texttt{string\_t}, \texttt{bool\_t}, \texttt{list\_t}, \texttt{dict\_t},
\texttt{obj\_t}) that serve documentation and constructor generation purposes without
runtime type enforcement. Fields support a \texttt{-static} modifier for class-level
storage and \texttt{public}/\texttt{private} visibility blocks. Every field has a default
value. VOO automatically generates three constructor variants: positional (\texttt{new}),
no-argument (\texttt{new()}), and named-argument (\texttt{new.args}). Custom constructors
can be declared for specialized initialization logic, and they are ordinary procedures
returning list values.

\begin{lstlisting}[language=Tcl, caption={VOO class declaration with multiple constructor variants}]
voo::class Person {
    public {
        string_t name "unknown"
        int_t age 0
        double_t salary 50000.0
    }

    method greet {} {
        return "Hello, I'm [get.name $this]"
    }
}

set p1 [Person::new "Alice" 30 75000.0]        ;# Positional
set p2 [Person::new()]                         ;# Defaults
set p3 [Person::new.args -name "Bob" -age 35]  ;# Named
\end{lstlisting}

\subsection{Accessors: Get, Set, Update}

For each field, VOO generates three accessor types. Getters receive the object by value
and return the field via \texttt{lindex}. Setters receive the variable name and modify
in-place via \texttt{upvar}/\texttt{lset}, providing copy-on-write safety. Updaters
temporarily detach a field into a local variable during a user-supplied script, preventing
COW propagation to the entire object during nested modifications---the
\texttt{try}/\texttt{finally} construct guarantees field reattachment even on exceptions.
Static fields use the \texttt{class.get.}/\texttt{class.set.} prefix convention.

\subsection{Methods and Inheritance}

Methods are procedures declared via \texttt{method} with optional modifiers:
none (\texttt{\$this} by value), \texttt{-static} (no \texttt{this}),
\texttt{-upvar} (\texttt{this} by reference), \texttt{-update \{fields\}} (fields
detached during body), and \texttt{-override} (validates parent method exists).
VOO supports single inheritance through \texttt{-extends}, where child classes inherit
parent fields with sequential indices and parent accessors are automatically available.
Parent methods can be imported explicitly via \texttt{importMethods}. Multiple inheritance
is excluded by design.

\subsection{Virtual Polymorphism}

VOO supports runtime polymorphic dispatch through the \texttt{-virtual} class flag and
\texttt{-virtual} method flag. A virtual class stores the concrete class namespace name
at index~0 of every instance---all other field indices shift up by one. This tag is an
interned \texttt{Tcl\_Obj*} shared by all instances of the same class; copy-on-write
ensures it is never duplicated on object copy, and its embedding as a literal at
class-definition time keeps constructor cost identical to non-virtual classes.

\begin{lstlisting}[language=Tcl, caption={Virtual polymorphic dispatch example}]
voo::class Shape -virtual {
    method area -virtual {} { return 0.0 }
}
voo::class Circle -extends Shape {
    public { double_t radius 1.0 }
    method area -override {} {
        return [expr {3.14159 * [get.radius $this] ** 2}]
    }
}
set s [Circle::new 5.0]
puts [Shape::area $s] ;# dispatches to Circle::area -> 78.54
\end{lstlisting}

For each \texttt{-virtual} method, VOO generates a \texttt{base.<name>} proc holding the
original body for direct parent calls, and makes the method itself a dispatcher that reads
index~0 and routes to the concrete class, falling back to \texttt{base.<name>} otherwise.
Child classes inherit virtual status automatically; \texttt{-override} methods are
auto-promoted to dispatchers, ensuring correct dispatch through deep inheritance chains.

\subsection{Visibility}

Fields and methods in \texttt{private \{ \}} blocks receive a \texttt{my.}~prefix and are
not exported. This naming-convention approach provides documentation-level protection
consistent with Tcl's philosophy---Itcl's enforcement mechanisms were bypassable through
namespace manipulation anyway.

\section{Implementation}
\label{sec:impl}

\subsection{Class Declaration Processing}

When \texttt{voo::class} is evaluated, VOO creates a namespace, parses field and method
declarations, assigns sequential field indices, and generates all accessors and
constructors. For inherited classes, parent field indices and defaults are copied first,
and new fields continue sequentially. According to experiments, declaring 1,000 small classes (5 fields, single method) using VOO costs approximately 300~\textmu s
versus 50~\textmu s for TclOO---an acceptable one-time cost during application
initialization, given the substantial runtime benefits.

\subsection{Copy-on-Write Optimization}

VOO's updater pattern avoids COW propagation during nested modifications by temporarily
extracting the field value into a local variable, setting the field slot to empty
(detaching it), executing the user's modification script, and reattaching the value in
a \texttt{finally} block. When the local variable holds the sole reference, modifications
proceed without copying the parent object.

\subsection{Migration Path to C++}

VOO's architecture enables seamless migration to compiled C++ via a companion
header-only template framework organized around three responsibilities.
\texttt{tcl::Type<T>} registers a custom \texttt{Tcl\_ObjType} for any C++ class,
providing COW semantics and reference counting through standard callbacks; only
\texttt{TypeName()} must be specialized. \texttt{tcl::obj\_cast} provides bidirectional
type casting between \texttt{Tcl\_Obj*} and C++ types with built-in specializations for
fundamental types. \texttt{TCLCPPG\_*} macros generate \texttt{Tcl\_CreateObjCommand}
registrations with lambda wrappers, argument validation, automatic type conversion, and
error handling.

The migration follows four steps: define a normal C++ class, specialize
\texttt{tcl::Type<T>}, register commands using macros, and use from Tcl---with the call
site identical to the VOO Tcl API\@. Migration requires no changes to caller code; only
the class declaration is replaced and commands re-registered by the C++ package
initializer.

\section{Evaluation}

\subsection{Experimental Setup}

All benchmarks ran on a Dual-Core Intel Xeon Gold 6240 CPU under Tcl~8.6.13 (stable)
and Tcl~9.0 (latest development). The evaluated frameworks are VOO (proposed), VOO~C++
(the benchmark \texttt{Point} class ported to a compiled C++ shared library via VOO's
template framework), TclOO (built-in), and Itcl~4.2. Time benchmarks use Tcl's
\texttt{time} command with 1,000 iterations (first execution discarded). Memory
benchmarks instantiate 100,000 objects in isolated \texttt{tclsh} processes and record resident set size
(RES) via \texttt{htop}. The test class is \texttt{Point} with 5 fields (two doubles, one
string, one integer, one boolean).

Virtual class overhead was also measured using an equivalent \texttt{VooVirtualPointImpl}
class (a concrete subclass of a \texttt{-virtual} base). No significant impact on object
creation time or memory footprint was found compared to non-virtual VOO classes. Over
100,000 objects, creation time increased by only $\sim$7\% on Tcl~8.6.13 (157~ms
vs.\ 147~ms) and $\sim$4\% on Tcl~9.0 (177~ms vs.\ 170~ms). Memory overhead was
negligible: $+$0.11~MB on Tcl~8.6.13 and $+$0.06~MB on Tcl~9.0. This near-zero memory
delta is explained by Tcl's reference-counting model: the namespace tag string stored at
index~0 of each virtual object is a single interned \texttt{Tcl\_Obj*} shared by all
instances of the same class; copy-on-write ensures it is never duplicated on object copy.
Virtual method dispatch carries $\sim$2~\textmu s overhead per call from the index-0 read
and conditional namespace routing, an upper bound incurred only on cross-class polymorphic
calls.

\subsection{Object Creation Performance}

\subsubsection*{Tcl 8.6.13}

\begin{table}[H]
\centering
\caption{Object creation performance on Tcl~8.6.13}
\begin{tabular}{llll}
\toprule
\textbf{Framework} & \textbf{Explicit (\textmu s)} & \textbf{Default (\textmu s)} & \textbf{Relative to VOO} \\
\midrule
VOO       & 0.414   & 0.397   & 1.00$\times$ (baseline) \\
VOO C++   & 0.565   & 0.334   & 1.36$\times$ slower / 1.19$\times$ faster \\
TclOO     & 3.536   & 2.972   & \textbf{8.5$\times$ / 7.5$\times$ slower} \\
Itcl      & 26.222  & 25.734  & \textbf{63$\times$ / 65$\times$ slower} \\
\bottomrule
\end{tabular}
\end{table}

\subsubsection*{Tcl 9.0}

\begin{table}[H]
\centering
\caption{Object creation performance on Tcl~9.0}
\begin{tabular}{llll}
\toprule
\textbf{Framework} & \textbf{Explicit (\textmu s)} & \textbf{Default (\textmu s)} & \textbf{Relative to VOO} \\
\midrule
VOO       & 0.525   & 0.550   & 1.00$\times$ (baseline) \\
VOO C++   & 0.576   & 0.458   & 1.10$\times$ slower / 1.20$\times$ faster \\
TclOO     & 9.246   & 3.829   & \textbf{17.6$\times$ / 7.0$\times$ slower} \\
Itcl      & 27.808  & 27.444  & \textbf{53$\times$ / 50$\times$ slower} \\
\bottomrule
\end{tabular}
\end{table}

VOO achieves direct list allocation with minimal overhead. VOO~C++ is comparable for
explicit creation and faster for defaults. TclOO's explicit constructor overhead increased
dramatically in Tcl~9.0 ($+$161\%), while Itcl remained consistently slow.

\subsection{Field Access Performance}

\subsubsection*{Tcl 8.6.13}

\begin{table}[H]
\centering
\caption{Field access performance on Tcl~8.6.13}
\begin{tabular}{lllll}
\toprule
\textbf{Operation} & \textbf{VOO (\textmu s)} & \textbf{VOO C++ (\textmu s)} & \textbf{TclOO (\textmu s)} & \textbf{Itcl (\textmu s)} \\
\midrule
Getter & 0.393 & 0.298 & 0.502 & 1.153 \\
Setter & 0.799 & 0.306 & 0.563 & 1.169 \\
\bottomrule
\end{tabular}
\end{table}

\subsubsection*{Tcl 9.0}

\begin{table}[H]
\centering
\caption{Field access performance on Tcl~9.0}
\begin{tabular}{lllll}
\toprule
\textbf{Operation} & \textbf{VOO (\textmu s)} & \textbf{VOO C++ (\textmu s)} & \textbf{TclOO (\textmu s)} & \textbf{Itcl (\textmu s)} \\
\midrule
Getter & 0.366 & 0.286 & 0.615 & 1.291 \\
Setter & 0.823 & 0.365 & 0.613 & 1.102 \\
\bottomrule
\end{tabular}
\end{table}

VOO~C++ delivers the fastest getters ($\sim$1.3$\times$ faster than VOO Tcl) and setters
(2.3--2.6$\times$ faster) across all frameworks. TclOO setters are 1.3--1.4$\times$
faster than VOO Tcl setters due to lighter variable-lookup overhead, but slower than
VOO~C++.

\subsection{Memory and Time Efficiency (100,000 Objects)}

\subsubsection*{Tcl 8.6.13}

\begin{table}[H]
\centering
\caption{Memory and time for 100,000 objects on Tcl~8.6.13}
\begin{tabular}{@{}llllll@{}}
\toprule
\textbf{Framework} & \textbf{Time} & \textbf{Memory} & \textbf{Bytes/Obj} & \textbf{vs.\ VOO (Time)} & \textbf{vs.\ VOO (Mem)} \\
\midrule
VOO       & 147~ms    & 58~MB   & $\sim$580   & 1.00$\times$             & 1.00$\times$ \\
VOO C++   & 149~ms    & 38~MB   & $\sim$380   & 1.01$\times$ slower      & \textbf{1.53$\times$ lighter} \\
TclOO     & 466~ms    & 257~MB  & $\sim$2,570 & \textbf{3.2$\times$ slower} & \textbf{4.4$\times$ heavier} \\
Itcl      & 2,781~ms  & 882~MB  & $\sim$8,820 & \textbf{18.9$\times$ slower} & \textbf{15.2$\times$ heavier} \\
\bottomrule
\end{tabular}
\end{table}

\subsubsection*{Tcl 9.0}

\begin{table}[H]
\centering
\caption{Memory and time for 100,000 objects on Tcl~9.0}
\begin{tabular}{@{}llllll@{}}
\toprule
\textbf{Framework} & \textbf{Time} & \textbf{Memory} & \textbf{Bytes/Obj} & \textbf{vs.\ VOO (Time)} & \textbf{vs.\ VOO (Mem)} \\
\midrule
VOO       & 170~ms    & 66.5~MB   & $\sim$665    & 1.00$\times$              & 1.00$\times$ \\
VOO C++   & 169~ms    & 40.3~MB   & $\sim$403    & \textbf{1.01$\times$ faster} & \textbf{1.65$\times$ lighter} \\
TclOO     & 577~ms    & 395~MB    & $\sim$3,950  & \textbf{3.4$\times$ slower}  & \textbf{5.9$\times$ heavier} \\
Itcl      & 3,014~ms  & 1,118~MB  & $\sim$11,180 & \textbf{17.7$\times$ slower} & \textbf{16.8$\times$ heavier} \\
\bottomrule
\end{tabular}
\end{table}

In a real-world scenario of creating 100,000 structured objects, VOO and VOO~C++ are
virtually identical in total wall-clock time ($\sim$147--170~ms), while TclOO takes
3.2--3.4$\times$ longer and Itcl is 18--19$\times$ slower. The memory picture is even
more pronounced: VOO~C++ holds the smallest footprint across all configurations
(6.8$\times$ lighter than TclOO on Tcl~8.6.13, widening to 9.8$\times$ on Tcl~9.0).
For million-object applications, VOO projects $\sim$665~MB versus TclOO's $\sim$3.95~GB
on Tcl~9.0.

\subsection{Cross-Version Analysis}

\begin{table}[H]
\centering
\caption{Performance change from Tcl~8.6.13 to Tcl~9.0}
\begin{tabular}{lllll}
\toprule
\textbf{Operation} & \textbf{VOO $\Delta$} & \textbf{VOO C++ $\Delta$} & \textbf{TclOO $\Delta$} & \textbf{Itcl $\Delta$} \\
\midrule
Creation (Explicit) & $+$26.8\% & $+$1.9\% & \textbf{$+$161.5\%} & $+$6.1\% \\
Creation (Default)  & $+$38.5\% & $+$37.1\% & $+$28.8\%           & $+$6.6\% \\
Getter              & $-$6.9\%  & $-$4.0\%  & $+$22.5\%           & $+$12.0\% \\
Setter              & $+$3.0\%  & $+$19.3\% & $+$8.9\%            & $-$5.7\% \\
\bottomrule
\end{tabular}
\end{table}

\begin{table}[H]
\centering
\caption{Memory change from Tcl~8.6.13 to Tcl~9.0}
\begin{tabular}{llll}
\toprule
\textbf{Framework} & \textbf{Memory (8.6)} & \textbf{Memory (9.0)} & \textbf{Change} \\
\midrule
VOO       & 58~MB    & 66.5~MB   & $+$14.7\% \\
VOO C++   & 38~MB    & 40.3~MB   & \textbf{$+$6.1\%} \\
TclOO     & 257~MB   & 395~MB    & $+$53.7\% \\
Itcl      & 882~MB   & 1,118~MB  & $+$26.8\% \\
\bottomrule
\end{tabular}
\end{table}

VOO~C++ is virtually immune to interpreter changes ($+$1.9\% creation, $+$6.1\% memory),
making it the most stable framework across versions. TclOO's 161.5\% explicit-construction
regression and 53.7\% memory increase in Tcl~9.0 are by far the largest among all
frameworks, improving VOO's relative advantage from 4.4$\times$ to 5.9$\times$ in memory
alone. Both VOO and VOO~C++ show improved getter performance in Tcl~9.0, indicating
bytecode optimizations benefit simple \texttt{lindex}/pointer-access operations. These
results validate the ``simplicity through composition'' philosophy: frameworks built from
native primitives scale better across language versions than those introducing new
infrastructure.

\section{Discussion}

\subsection{Application Domains}

VOO is recommended for high-performance data backends managing hundreds of thousands to
millions of structured objects, long-running processes where manual destructor management
introduces maintenance burden, C++ migration candidates, and new development projects
prioritizing performance and clarity. TclOO or Itcl may be preferable for existing stable
codebases where migration costs outweigh benefits, teams preferring traditional
\texttt{\$obj method} syntax, or applications requiring multiple inheritance or TclOO's
mixin system.

\subsection{Comparison Summary}

\begin{table}[H]
\centering
\caption{Feature and performance comparison across OO frameworks}
\begin{tabular}{L{3.2cm}L{2.5cm}L{2.5cm}L{2.2cm}L{2.2cm}}
\toprule
\textbf{Aspect} & \textbf{VOO (Tcl)} & \textbf{VOO (C++)} & \textbf{TclOO} & \textbf{Itcl} \\
\midrule
Object Creation   & \textbf{7--18$\times$ faster} & \textbf{6--16$\times$ faster} & Baseline & Slowest \\
Memory/Object     & \textbf{4--6$\times$ lighter} & \textbf{7--10$\times$ lighter} & Baseline & Heaviest \\
Getters           & \textbf{Fastest (Tcl)} & \textbf{Fastest overall} & Good & Slow \\
Setters           & Good & \textbf{Fastest overall} & Faster than VOO Tcl & Slow \\
Destructor        & \textbf{None (auto)} & \textbf{None (auto)} & Required & Required \\
Inheritance       & Single & Single & Multiple & Multiple \\
C++ Migration     & \textbf{Seamless} & \textbf{Already C++} & Redesign needed & Redesign needed \\
Cross-Version     & \textbf{Excellent} & \textbf{Best} & Moderate & Poor \\
\bottomrule
\end{tabular}
\end{table}

\subsection{Philosophical Alignment}

VOO embodies the ``Tao of Tcl''~\cite{ousterhout1994}: objects are lists, classes are
namespaces, accessor patterns are explicit and predictable, object behavior matches strings
and lists, and all objects maintain human-readable string representations. While TclOO
improved upon Itcl's inconsistencies, it retains reference semantics and procedure handle
mechanisms that diverge from Tcl's native type behavior.

\subsection{Limitations and Future Work}

Current limitations include single inheritance only (by design), no mixin support, and
class declaration overhead approximately 4--6$\times$ slower than TclOO (an acceptable
one-time cost). Future directions include syntax exploration, automated TclOO $\to$ VOO
migration tools, pattern libraries, and community feedback integration.

\section{Related Work}

\begin{table}[H]
\centering
\caption{Object models in scripting languages}
\begin{tabular}{llll}
\toprule
\textbf{Language}   & \textbf{Object Model}            & \textbf{Semantics}                          & \textbf{Memory Mgmt} \\
\midrule
Python              & \texttt{\_\_dict\_\_}            & Reference                                   & GC \\
Ruby                & Everything is object             & Reference                                   & GC \\
JavaScript          & Prototype-based                  & Reference                                   & GC \\
Lua                 & Tables + metatables              & Reference                                   & GC \\
\textbf{VOO}        & Lists + namespaces               & \textbf{Value} / Ref.\ by variable name     & \textbf{COW + Refcount} \\
\bottomrule
\end{tabular}
\end{table}

\begin{table}[H]
\centering
\caption{Tcl OO frameworks comparison}
\begin{tabular}{lllll}
\toprule
\textbf{Framework}         & \textbf{Basis}      & \textbf{Semantics}                      & \textbf{Status} \\
\midrule
Itcl~\cite{dejong,tclwiki-itcl}  & Extension     & Reference                               & Legacy \\
TclOO~\cite{fellows2006,tcloo}   & Core (8.6+)   & Reference                               & Standard \\
Snit                              & Pure Tcl      & Reference                               & Widgets \\
XOTcl~\cite{neumann2000}          & Extension     & Reference                               & Research \\
\textbf{VOO}                      & Native types  & \textbf{Value} / Ref.\ by variable name & \textbf{This work} \\
\bottomrule
\end{tabular}
\end{table}

Value semantics are gaining recognition in modern systems languages---C++11 move
semantics, Rust's ownership model, and Swift's value types. VOO demonstrates that value
semantics and object orientation are complementary rather than mutually exclusive,
extending these benefits to Tcl scripting.

\section{Conclusions}

I have presented Vanilla Object Orientation (VOO), a high-performance OO framework for
Tcl that constructs classes from native data structures rather than introducing additional
abstraction layers. Objects are plain lists with automatic memory management through
Tcl's existing mechanisms, achieving 7--18$\times$ faster object creation and
4--6$\times$ superior memory efficiency versus TclOO across Tcl~8.6.13 and~9.0.
VOO~C++ extends these gains further with the fastest field access and smallest memory
footprint of all evaluated frameworks. Cross-version analysis shows that VOO's
compositional design scales better than framework-based approaches---TclOO's memory
footprint increased 53.7\% in Tcl~9.0 while VOO increased only 14.8\% and VOO~C++ just
6.1\%.

The evolution from Itcl (1990s) $\to$ TclOO (2013) $\to$ VOO represents the Tcl
community's progressive understanding that OO support should harmonize with the
language's foundational philosophy. VOO provides organizational patterns using Tcl's
established primitives, with an incremental migration path (lists $\to$ VOO $\to$ C++)
that preserves API compatibility throughout. Our results suggest that philosophical
alignment with core language principles provides not only conceptual elegance but also
practical longevity: simplicity enables performance, and frameworks built from native
primitives scale better across language versions than those introducing new
infrastructure.

\paragraph{Availability.}
VOO is available as open-source software. License: MIT\@.

Initial version of the code and supplementary material is provided in the appendices:
\begin{sloppypar}
\begin{itemize}
  \item \textbf{Appendix~\ref{app:quickref}} --- Quick Reference (class declaration
        syntax, field types, acessors, constructors)
  \item \textbf{Appendix~\ref{app:benchmark}} --- Benchmark Methodology (timing
        procedure, full test class implementations, memory measurement)
  \item \textbf{Appendix~\ref{app:syntax}} --- VOO Syntax Sugar vs.\ Raw Vanilla Tcl
        (side-by-side code generation walkthrough)
  \item \textbf{Appendix~\ref{app:cpp}} --- C++/Tcl Glue Code Template Framework
        (complete source for \texttt{tcl/type.h}, \texttt{tcl/obj\_cast.h},
        \texttt{tcl/cpp\_api.h})
  \item \textbf{Appendix~\ref{app:voopkg}} --- VOO Tcl Package (complete source of \texttt{voo.tcl})
\end{itemize}
\end{sloppypar}

\section*{Acknowledgments}
\addcontentsline{toc}{section}{Acknowledgments}

I thank Donal Fellows for TclOO (TIP~\#257)~\cite{fellows2006}, and the
Itcl~\cite{dejong,tclwiki-itcl} and XOTcl~\cite{neumann2000} maintainers who first
demonstrated OO in Tcl. At Cadence Design Systems, special thanks to everyone who
contributed in any way to this topic, especially Francesco Lertora, Elaine Scartezzini,
and Tiago Alves. I acknowledge John Ousterhout~\cite{ousterhout1994} and the broader
Tcl community for over 30 years of maintaining Tcl's philosophical coherence.

The original ideas, concepts, design, implementation, and experiments presented in this
work are the author's own. Generative AI tools were used to refine
content clarity, formatting, and presentation. Generative AI was also employed to assist
in generating code in accordance with the author's original concepts and specifications.

\appendix
\section{Quick Reference}
\label{app:quickref}

This appendix provides concise reference documentation for VOO's primary constructs and
usage patterns.

\subsection*{Class Declaration}

\begin{lstlisting}[language=Tcl, caption={VOO class declaration template}]
voo::class ClassName ?-virtual? ?-extends ParentClass? {
    public {
        type_t fieldName defaultValue
        type_t -static staticFieldName defaultValue

        method methodName {args} { body }
        method methodName {args} -static { body }
        method methodName {args} -upvar { body }
        method methodName {args} -update {fields} { body }
        method methodName {args} -override { body }
        method methodName -virtual {args} { body }
    }

    private {
        type_t fieldName defaultValue
        method methodName {args} { body }
    }

    constructor {args} { return [list ...] }

    importMethods {parentMethod1 parentMethod2}
}
\end{lstlisting}

\subsection*{Field Types}

\begin{table}[H]
\centering
\caption{VOO field type annotations}
\begin{tabular}{lll}
\toprule
\textbf{Type} & \textbf{Description} & \textbf{Example} \\
\midrule
\texttt{double\_t} & Floating-point & \texttt{double\_t x 0.0} \\
\texttt{int\_t}    & Integer        & \texttt{int\_t count 0} \\
\texttt{string\_t} & String         & \texttt{string\_t name ""} \\
\texttt{bool\_t}   & Boolean        & \texttt{bool\_t active 1} \\
\texttt{list\_t}   & List           & \texttt{list\_t items [list]} \\
\texttt{dict\_t}   & Dictionary     & \texttt{dict\_t data [dict create]} \\
\texttt{obj\_t}    & Any object     & \texttt{obj\_t nested \{\}} \\
\bottomrule
\end{tabular}
\end{table}

\subsection*{Accessors}

\begin{lstlisting}[language=Tcl, caption={Accessor patterns}]
# Instance field accessors
set val [Class::get.field $obj]
Class::set.field obj value
Class::update.field obj temp { modify $temp }

# Static field accessors
set val [Class::class.get.staticField]
Class::class.set.staticField value
\end{lstlisting}

\subsection*{Constructors}

\begin{lstlisting}[language=Tcl, caption={Three constructor variants}]
set obj [Class::new arg1 arg2 ...]       ;# Positional
set obj [Class::new()]                    ;# Default values
set obj [Class::new.args -field1 val ...] ;# Named arguments
\end{lstlisting}

\subsection*{Virtual Dispatch}

\begin{lstlisting}[language=Tcl, caption={Virtual dispatch patterns}]
# Call through abstract base --- dispatches to concrete class at runtime
set result [BaseClass::virtualMethod $obj]

# Direct parent-body call from within an override
BaseClass::base.virtualMethod $this
\end{lstlisting}

\section{Benchmark Methodology}
\label{app:benchmark}

This appendix describes the experimental procedures used for the performance evaluation.

\subsection*{Timing Methodology}

All time measurements use Tcl's built-in \texttt{time} command, which executes a script
for a specified number of iterations and returns the average execution time per iteration
in microseconds. To ensure fair comparisons, a \texttt{profile} wrapper discards the first
execution (allowing Tcl's bytecode compiler to optimize the procedure) before measuring:

\begin{lstlisting}[language=Tcl, caption={Timing wrapper discarding the first run}]
proc profile {body times} {
    uplevel 1 $body; # discard first run to compile tcl procs
    uplevel 1 [list time $body $times]
}
\end{lstlisting}

Each benchmark invocation runs with 1,000 iterations for object creation, getter, and
setter tests. Class declaration benchmarks also use 1,000 iterations but measure the time
to declare complete classes with multiple fields and methods.

\subsection*{Test Classes}

The benchmark suite uses a \texttt{Point} class with five fields of mixed types to
represent realistic object complexity:

\begin{lstlisting}[language=Tcl, caption={VOO \texttt{Point} implementation}]
voo::class VooPoint {
    public {
        double_t x 0.0
        double_t y 0.0
        string_t name "point"
        int_t id 0
        bool_t active 1
    }

    method distance {} {
        set dx [get.x $this]
        set dy [get.y $this]
        return [expr {sqrt($dx * $dx + $dy * $dy)}]
    }
}
\end{lstlisting}

\begin{lstlisting}[language=Tcl, caption={TclOO \texttt{Point} implementation}]
oo::class create TclooPoint {
    variable x 0.0
    variable y 0.0
    variable name "point"
    variable id 0
    variable active 1

    constructor {x_ y_ name_ id_ active_} {
        set x $x_; set y $y_; set name $name_; set id $id_; set active $active_
    }

    method getX {} { return $x }
    method setX {value} { set x $value }
    method distance {} { return [expr {sqrt($x * $x + $y * $y)}] }
}
\end{lstlisting}

\begin{lstlisting}[language=Tcl, caption={Itcl \texttt{Point} implementation}]
itcl::class ItclPoint {
    public variable x 0.0
    public variable y 0.0
    public variable name "point"
    public variable id 0
    public variable active 1

    constructor {x_ y_ name_ id_ active_} {
        set x $x_; set y $y_; set name $name_; set id $id_; set active $active_
    }

    method getX {} { return $x }
    method setX {value} { set x $value }
    method distance {} { return [expr {sqrt($x * $x + $y * $y)}] }
}
\end{lstlisting}

The VOO~C++ implementation is loaded from a pre-compiled shared library:

\begin{lstlisting}[language=Tcl, caption={Loading and using the VOO C++ Point class}]
# VOO C++ --- load compiled shared library built from point.cpp
load <path/to/libvoopoint_cpp.so> Point

set obj [CppVooPoint::new 1.0 2.0 "test" 1 1]
set obj [CppVooPoint::new()]                    ;# default constructor
set x   [CppVooPoint::get.x $obj]              ;# getter --- object by value
CppVooPoint::set.x obj 3.14                    ;# setter --- variable by name
\end{lstlisting}

\noindent\textbf{Note on Class Declaration Performance:} VOO~C++ class declarations
happen entirely at C++ compile time; there is no runtime class registration cost to
measure. Package loading time (the one-time cost of the \texttt{load} command) is excluded
from all benchmark comparisons because it is an initialization step analogous to
\texttt{source}-ing the VOO Tcl framework or loading Itcl.

\subsection*{Memory Measurement}

Memory measurements were performed separately using a dedicated memory benchmark script.
Each scenario runs in an isolated \texttt{tclsh} process. The process resident set size
(RES) is recorded via \texttt{htop} after creating 100,000 objects of each class type.

\subsection*{Benchmark Categories}

The benchmark suite measures five distinct performance metrics:
\begin{enumerate}
  \item \textbf{Object Creation (Explicit Values):} Instantiate objects with all field
        values explicitly provided to the constructor.
  \item \textbf{Object Creation (Default Values):} Instantiate objects using the
        no-argument constructor with all default field values.
  \item \textbf{Setter Performance:} Modify a single field value through the setter.
  \item \textbf{Getter Performance:} Retrieve a single field value through the getter.
  \item \textbf{Class Declaration Performance:} Declare a complete class with fields and
        methods (one-time initialization cost).
\end{enumerate}

\section{VOO Syntax Sugar vs.\ Raw Vanilla Tcl}
\label{app:syntax}

This appendix demonstrates the value of VOO's syntax sugar by comparing a \texttt{Point}
class written with VOO's declarative syntax against the equivalent raw Tcl that VOO
generates internally.

\subsection*{C.1 Class Declaration and Field Indices}

\begin{lstlisting}[language=Tcl, caption={VOO syntax sugar: class and fields}]
voo::class Point {
    public {
        double_t x 0.0
        double_t y 0.0
        string_t name "point"
    }
}
\end{lstlisting}

\begin{lstlisting}[language=Tcl, caption={Raw Vanilla Tcl generated by VOO}]
namespace eval Point {
    variable x 0
    variable y 1
    variable name 2

    variable __defaultObj [list 0.0 0.0 "point"]
    variable __fields [list x y name]
}
\end{lstlisting}

\subsection*{C.2 Constructors}

\subsubsection*{C.2.1 Positional Constructor}
\begin{lstlisting}[language=Tcl]
# VOO: auto-generated Point::new
set p [Point::new 1.5 2.5 "A"]

# Raw (generated):
proc Point::new {x y name} { return [list $x $y $name] }
\end{lstlisting}

\subsubsection*{C.2.2 No-Argument Constructor}
\begin{lstlisting}[language=Tcl]
# VOO:
set p [Point::new()]

# Raw (generated):
proc Point::new() {} {
    variable __defaultObj
    return $__defaultObj
}
\end{lstlisting}

\subsubsection*{C.2.3 Named-Argument Constructor}
\begin{lstlisting}[language=Tcl]
# VOO:
set p [Point::new.args -x 1.5 -y 2.5]

# Raw (generated):
proc Point::new.args {args} {
    variable __defaultObj
    set obj $__defaultObj
    if {[catch {dict size $args}]} {
        error "Constructor argument must be a list of '-<field> <value>' pairs"
    }
    dict for {key value} $args {
        if {[string index $key 0] ne "-"} {
            error "Constructor argument keys must start with '-', got '$key'"
        }
        set field [string range $key 1 end]
        set setter set.$field
        if {[info commands $setter] ne ""} {
            $setter obj $value
        } else {
            set setter my.set.$field
            if {[info commands $setter] ne ""} {
                $setter obj $value
            } else {
                error "Unknown field option: $field"
            }
        }
    }
    return $obj
}
\end{lstlisting}

\subsection*{C.3 Getters}
\begin{lstlisting}[language=Tcl]
# VOO: auto-generated
set x [Point::get.x $p]

# Raw (generated per field):
proc Point::get.x {this} { variable x; return [lindex $this $x] }
proc Point::get.y {this} { variable y; return [lindex $this $y] }
proc Point::get.name {this} { variable name; return [lindex $this $name] }
\end{lstlisting}

\subsection*{C.4 Setters}
\begin{lstlisting}[language=Tcl]
# VOO: auto-generated
Point::set.x p 3.14

# Raw (generated per field):
proc Point::set.x {thisVar value} {
    variable x
    upvar $thisVar this
    lset this $x $value
}
\end{lstlisting}

Setters use \texttt{upvar} to receive the variable name, enabling in-place modification
with copy-on-write safety.

\subsection*{C.5 Updaters (Copy-on-Write Optimization)}
\begin{lstlisting}[language=Tcl]
# VOO: auto-generated
Point::update.x p temp { set temp [expr {$temp * 2}] }

# Raw (generated per field):
proc Point::update.x {thisVar tempVar body} {
    variable x
    upvar $thisVar this
    upvar $tempVar temp
    try {
        set temp [lindex $this $x]
        lset this $x {}
        uplevel $body
    } finally {
        lset this $x $temp
    }
}
\end{lstlisting}

\subsection*{C.6 Instance Methods}
\begin{lstlisting}[language=Tcl]
# VOO:
voo::class Point {
    method distance {} {
        set dx [get.x $this]; set dy [get.y $this]
        return [expr {sqrt($dx*$dx + $dy*$dy)}]
    }
}

# Raw (generated):
proc Point::distance {this} {
    variable x; variable y
    set dx [lindex $this $x]; set dy [lindex $this $y]
    return [expr {sqrt($dx*$dx + $dy*$dy)}]
}
\end{lstlisting}

\subsection*{C.7 Static Fields}
\begin{lstlisting}[language=Tcl]
# VOO:
voo::class Point {
    public { int_t -static count 0 }
}
set n [Point::class.get.count]

# Raw (generated):
namespace eval Point {
    variable count 0
    proc class.get.count {} { variable count; return $count }
    proc class.set.count {value} { variable count; set count $value }
}
\end{lstlisting}

\subsection*{C.8 Virtual Classes and Methods}

\subsubsection*{C.8.1 Virtual Base Class Declaration}
\begin{lstlisting}[language=Tcl, caption={VOO virtual base}]
voo::class Shape -virtual {
    public { double_t radius 1.0 }
    method area -virtual {} { return 0.0 }
}
\end{lstlisting}

\begin{lstlisting}[language=Tcl, caption={Raw Vanilla Tcl generated for virtual base}]
namespace eval Shape {
    # Index 0 is permanently reserved for the class namespace tag.
    variable radius 1          ;# field index = 1
    variable __defaultObj [list ::Shape 1.0]
    variable __fields [list radius]
    variable __voo_is_virtual_class 1

    proc new {radius} { return [list ::Shape $radius] }

    # base.area holds the original body for direct parent calls
    proc base.area {this} { return 0.0 }

    # area is a dispatcher
    proc area {this} {
        set __voo_cls [lindex $this 0]
        if {$__voo_cls ne [namespace current] && \
            [info commands ${__voo_cls}::area] ne {}} {
            return [${__voo_cls}::area $this]
        }
        return [base.area $this]
    }
}
\end{lstlisting}

\subsubsection*{C.8.2 Virtual Child Class Declaration}
\begin{lstlisting}[language=Tcl, caption={Child class overriding a virtual method}]
voo::class Circle -extends Shape {
    method area -override {} {
        return [expr {3.14159 * [get.radius $this] ** 2}]
    }
}
\end{lstlisting}

\begin{lstlisting}[language=Tcl, caption={Raw Vanilla Tcl generated for child class}]
namespace eval Circle {
    variable radius 1          ;# same index as Shape::radius
    variable __defaultObj [list ::Circle 1.0]
    variable __voo_is_virtual_class 1

    proc new {radius} { return [list ::Circle $radius] }

    proc base.area {this} {
        return [expr {3.14159 * [get.radius $this] ** 2}]
    }
    proc area {this} {
        set __voo_cls [lindex $this 0]
        if {$__voo_cls ne [namespace current] && \
            [info commands ${__voo_cls}::area] ne {}} {
            return [${__voo_cls}::area $this]
        }
        return [base.area $this]
    }
}
\end{lstlisting}

\subsubsection*{C.8.3 Calling the Parent Body from an Override}
\begin{lstlisting}[language=Tcl]
voo::class ColoredCircle -extends Circle {
    public { string_t color "red" }
    method area -override {} {
        # Add 10% for visual padding
        set base [Circle::base.area $this]
        return [expr {$base * 1.1}]
    }
}
\end{lstlisting}

\section{C++/Tcl Glue Code Template Framework}
\label{app:cpp}

Complete source for the three-header template framework described in
Section~\ref{sec:impl}. Requires C++17 and Tcl~8.6 C~API headers.

\subsection*{D.1 \texttt{tcl/type.h} --- Custom \texttt{Tcl\_ObjType} Registration}

\begin{lstlisting}[language=C++, caption={\texttt{tcl/type.h} --- type registration template}]
#pragma once
#include <tcl.h>
#include <exception>
#include <string>
#include <sstream>
#include <cstring>

namespace tcl {

template <typename T>
class Type {
public:
    static Tcl_ObjType* GetType() {
        static std::string nameStorage(TypeName());
        static Tcl_ObjType type = {
            nameStorage.c_str(),
            FreeInternalRep,
            DupInternalRep,
            UpdateString,
            SetFromAny
        };
        static bool registered = false;
        if (!registered) { Tcl_RegisterObjType(&type); registered = true; }
        return &type;
    }

    static Tcl_Obj* New(T* value) {
        Tcl_Obj* obj = Tcl_NewObj();
        if (!obj) throw std::runtime_error("Failed to allocate Tcl object");
        obj->bytes = NULL;
        obj->typePtr = GetType();
        obj->internalRep.otherValuePtr = static_cast<void*>(value);
        return obj;
    }

    static void FreeInternalRep(Tcl_Obj* obj) {
        delete static_cast<T*>(obj->internalRep.otherValuePtr);
        obj->internalRep.otherValuePtr = nullptr;
    }

    static void DupInternalRep(Tcl_Obj* src, Tcl_Obj* dup) {
        dup->internalRep.otherValuePtr =
            new T(*static_cast<T*>(src->internalRep.otherValuePtr));
        dup->typePtr = GetType();
    }

    static void UpdateString(Tcl_Obj* obj) {
        T* value = static_cast<T*>(obj->internalRep.otherValuePtr);
        std::string str = ToString(*value);
        obj->bytes = Tcl_Alloc(str.size() + 1);
        strcpy(obj->bytes, str.c_str());
        obj->length = str.size();
    }

    static int SetFromAny(Tcl_Interp* interp, Tcl_Obj* obj) {
        try {
            T value = FromAny(interp, obj);
            if (obj->typePtr && obj->typePtr->freeIntRepProc)
                obj->typePtr->freeIntRepProc(obj);
            obj->internalRep.otherValuePtr = new T(value);
            obj->typePtr = GetType();
            Tcl_InvalidateStringRep(obj);
            return TCL_OK;
        } catch (const std::exception& e) {
            if (interp) Tcl_SetObjResult(interp, Tcl_NewStringObj(e.what(), -1));
            return TCL_ERROR;
        }
    }

    static T* GetInternalRep(Tcl_Interp* interp, Tcl_Obj* obj) {
        if (obj->typePtr != GetType()) {
            if (SetFromAny(interp, obj) != TCL_OK) {
                std::ostringstream oss;
                oss << "Failed to convert to \"" << TypeName() << "\"";
                throw std::runtime_error(oss.str());
            }
        }
        return static_cast<T*>(obj->internalRep.otherValuePtr);
    }

    // Default ToString --- throws; specialize to enable `puts $obj`
    static std::string ToString(const T&) {
        std::ostringstream oss;
        oss << "Type \"" << TypeName() << "\" can't be cast to string";
        throw std::runtime_error(oss.str());
    }

    // Default FromAny falls back to string conversion
    static T FromAny(Tcl_Interp*, Tcl_Obj* const obj) {
        return FromString(Tcl_GetString(obj));
    }

    static T FromString(const std::string&) {
        std::ostringstream oss;
        oss << "Type \"" << TypeName() << "\" can't be cast from string";
        throw std::runtime_error(oss.str());
    }

private:
    static const char* TypeName();  // Must be specialized per type
};

} // namespace tcl
\end{lstlisting}

\begin{table}[H]
\centering
\caption{Specialization points for \texttt{tcl::Type<T>}}
\begin{tabular}{lll}
\toprule
\textbf{Function} & \textbf{Required} & \textbf{Purpose} \\
\midrule
\texttt{TypeName()} & Yes & Unique name for Tcl's type registry \\
\texttt{ToString(const T\&)} & No & Serializes \texttt{T} for \texttt{puts \$obj} \\
\texttt{FromAny(interp, obj)} & No & Parses a \texttt{Tcl\_Obj} into \texttt{T} \\
\texttt{FromString(const string\&)} & No & Fallback used by default \texttt{FromAny} \\
\bottomrule
\end{tabular}
\end{table}

\subsection*{D.2 \texttt{tcl/obj\_cast.h} --- Bidirectional Type Casting}

\begin{lstlisting}[language=C++, caption={\texttt{tcl/obj\_cast.h} (excerpt showing key specializations)}]
#pragma once
#include <tcl.h>
#include <string>
#include <stdexcept>
#include <type_traits>
#include "tcl/type.h"

namespace tcl { namespace obj_cast {

    template <typename T>
    Tcl_Obj* from(const T&) {
        static_assert(!std::is_same_v<T, T>,
            "No tcl::obj_cast::from<T> specialization for this type.");
        return nullptr;
    }

    template <typename T>
    T to(Tcl_Interp* interp, Tcl_Obj* const obj) {
        if constexpr (std::is_pointer_v<T>)
            return tcl::Type<std::remove_pointer_t<T>>::GetInternalRep(interp, obj);
        else
            return *tcl::Type<T>::GetInternalRep(interp, obj);
    }

    // int
    template <> inline Tcl_Obj* from<int>(const int& v) { return Tcl_NewIntObj(v); }
    template <> inline int to<int>(Tcl_Interp* i, Tcl_Obj* const o) {
        int v; Tcl_GetIntFromObj(i, o, &v); return v; }

    // double
    template <> inline Tcl_Obj* from<double>(const double& v) { return Tcl_NewDoubleObj(v); }
    template <> inline double to<double>(Tcl_Interp* i, Tcl_Obj* const o) {
        double v; Tcl_GetDoubleFromObj(i, o, &v); return v; }

    // std::string
    template <> inline Tcl_Obj* from<std::string>(const std::string& v) {
        return Tcl_NewStringObj(v.c_str(), -1); }
    template <> inline std::string to<std::string>(Tcl_Interp*, Tcl_Obj* const o) {
        int len; const char* s = Tcl_GetStringFromObj(o, &len);
        return std::string(s, len); }

    // bool
    template <> inline Tcl_Obj* from<bool>(const bool& v) { return Tcl_NewBooleanObj(v); }
    template <> inline bool to<bool>(Tcl_Interp* i, Tcl_Obj* const o) {
        int v; Tcl_GetBooleanFromObj(i, o, &v); return static_cast<bool>(v); }

    // Tcl_Obj* passthrough
    template <> inline Tcl_Obj* from<Tcl_Obj*>(Tcl_Obj* const& v) { return v; }
    template <> inline Tcl_Obj* to<Tcl_Obj*>(Tcl_Interp*, Tcl_Obj* const o) { return o; }

}} // namespace tcl::obj_cast
\end{lstlisting}

\subsection*{D.3 \texttt{tcl/cpp\_api.h} --- Glue Macros and Template Functions}

\begin{lstlisting}[language=C++, caption={Key macros from \texttt{tcl/cpp\_api.h}}]
// Constructor: new T(arg1, arg2, ...)
#define TCLCPPG_CREATE_CMD(interp, cmdName, className, ...) \
    Tcl_CreateObjCommand(interp, cmdName, \
        [](ClientData cd, Tcl_Interp *ti, int oc, Tcl_Obj *const ov[]) -> int { \
            return tcl::cpp_api::create<className, __VA_ARGS__>(cd, ti, oc, ov); \
        }, NULL, NULL)

// Getter (no extra args): object by value
#define TCLCPPG_GETTER_CMD_NOARGS(interp, cmdName, className, returnType, methodName) \
    Tcl_CreateObjCommand(interp, cmdName, \
        [](ClientData cd, Tcl_Interp *ti, int oc, Tcl_Obj *const ov[]) -> int { \
            return tcl::cpp_api::getter<className, returnType>( \
                cd, ti, oc, ov, &className::methodName); \
        }, NULL, NULL)

// Setter (with extra args): object by variable name
#define TCLCPPG_SETTER_CMD(interp, cmdName, className, returnType, methodName, ...) \
    Tcl_CreateObjCommand(interp, cmdName, \
        [](ClientData cd, Tcl_Interp *ti, int oc, Tcl_Obj *const ov[]) -> int { \
            return tcl::cpp_api::setter<className, returnType, __VA_ARGS__>( \
                cd, ti, oc, ov, &className::methodName); \
        }, NULL, NULL)
\end{lstlisting}

\begin{table}[H]
\centering
\caption{Summary of \texttt{TCLCPPG\_*} macros}
\begin{tabular}{@{}L{5.8cm}L{3.0cm}L{5.2cm}@{}}
\toprule
\textbf{Macro} & \textbf{Tcl convention} & \textbf{C++ side} \\
\midrule
\texttt{TCLCPPG\_CREATE\_CMD} & \texttt{name arg1 arg2 ...} & \texttt{new Class(args)} \\
\texttt{TCLCPPG\_GETTER\_CMD\_NOARGS} & \texttt{name object} & \texttt{const} member, no extra args \\
\texttt{TCLCPPG\_GETTER\_CMD} & \texttt{name object arg1 ...} & \texttt{const} member with extra args \\
\texttt{TCLCPPG\_SETTER\_CMD\_NOARGS} & \texttt{name objectVar} & non-\texttt{const} member, no extra args \\
\texttt{TCLCPPG\_SETTER\_CMD} & \texttt{name objectVar arg1 ...} & non-\texttt{const} member with extra args \\
\texttt{TCLCPPG\_FREE\_CMD} & \texttt{name arg1 arg2 ...} & free function or static method \\
\bottomrule
\end{tabular}
\end{table}

\subsection*{D.4 Complete Example: \texttt{CppVooPoint} (\texttt{point.cpp})}

\begin{lstlisting}[language=C++, caption={Step 1--2: C++ class and \texttt{tcl::Type} specialization}]
#include <tcl.h>
#include <string>
#include <cmath>
#include <sstream>
#include "tcl/cpp_api.h"

class VooPoint {
    double m_x, m_y;
    std::string m_name;
    int m_id;
    bool m_active;
public:
    VooPoint(double x, double y, const std::string& name, int id, bool active)
        : m_x(x), m_y(y), m_name(name), m_id(id), m_active(active) {}
    VooPoint() : m_x(0.0), m_y(0.0), m_name("point"), m_id(0), m_active(true) {}

    double             getX()      const { return m_x; }
    double             getY()      const { return m_y; }
    const std::string& getName()   const { return m_name; }
    int                getId()     const { return m_id; }
    bool               getActive() const { return m_active; }
    void setX(double v)            { m_x = v; }
    void setY(double v)            { m_y = v; }
    void setName(std::string v)    { m_name = std::move(v); }
    void setId(int v)              { m_id = v; }
    void setActive(bool v)         { m_active = v; }
    double distance() const { return std::sqrt(m_x*m_x + m_y*m_y); }
};

namespace tcl {
template <> const char* Type<VooPoint>::TypeName() { return "VooPoint"; }
template <> std::string Type<VooPoint>::ToString(const VooPoint& p) {
    std::ostringstream oss;
    oss << p.getX() << " " << p.getY() << " "
        << p.getName() << " " << p.getId() << " " << p.getActive();
    return oss.str();
}
template <> VooPoint Type<VooPoint>::FromAny(Tcl_Interp* interp, Tcl_Obj* const obj) {
    int objc; Tcl_Obj** objv;
    if (Tcl_ListObjGetElements(interp, obj, &objc, &objv) != TCL_OK || objc != 5)
        throw std::runtime_error("Expected list of 5 elements: x y name id active");
    return VooPoint(
        obj_cast::to<double>(interp, objv[0]),
        obj_cast::to<double>(interp, objv[1]),
        obj_cast::to<std::string>(interp, objv[2]),
        obj_cast::to<int>(interp, objv[3]),
        obj_cast::to<bool>(interp, objv[4]));
}
}
\end{lstlisting}

\begin{lstlisting}[language=C++, caption={Step 3--4: default constructor helper and package initializer}]
// Zero-argument default constructor
static int VooPoint_newDefault(ClientData, Tcl_Interp* interp,
                               int objc, Tcl_Obj* const objv[]) {
    if (objc != 1) { Tcl_WrongNumArgs(interp, 1, objv, ""); return TCL_ERROR; }
    try {
        Tcl_SetObjResult(interp, tcl::Type<VooPoint>::New(new VooPoint()));
    } catch (const std::exception& e) {
        Tcl_SetObjResult(interp, Tcl_NewStringObj(e.what(), -1));
        return TCL_ERROR;
    }
    return TCL_OK;
}

extern "C" { int Point_Init(Tcl_Interp* interp); }

int Point_Init(Tcl_Interp* interp) {
    Tcl_CreateNamespace(interp, "::CppVooPoint", NULL, NULL);

    // Constructors
    TCLCPPG_CREATE_CMD(interp, "::CppVooPoint::new",
                       VooPoint, double, double, std::string, int, bool);
    Tcl_CreateObjCommand(interp, "::CppVooPoint::new()",
                         VooPoint_newDefault, NULL, NULL);

    // Getters
    TCLCPPG_GETTER_CMD_NOARGS(interp, "::CppVooPoint::get.x",     VooPoint, double,             getX);
    TCLCPPG_GETTER_CMD_NOARGS(interp, "::CppVooPoint::get.y",     VooPoint, double,             getY);
    TCLCPPG_GETTER_CMD_NOARGS(interp, "::CppVooPoint::get.name",  VooPoint, const std::string&, getName);
    TCLCPPG_GETTER_CMD_NOARGS(interp, "::CppVooPoint::get.id",    VooPoint, int,                getId);
    TCLCPPG_GETTER_CMD_NOARGS(interp, "::CppVooPoint::get.active",VooPoint, bool,               getActive);
    TCLCPPG_GETTER_CMD_NOARGS(interp, "::CppVooPoint::distance",  VooPoint, double,             distance);

    // Setters
    TCLCPPG_SETTER_CMD(interp, "::CppVooPoint::set.x",     VooPoint, void, setX,     double);
    TCLCPPG_SETTER_CMD(interp, "::CppVooPoint::set.y",     VooPoint, void, setY,     double);
    TCLCPPG_SETTER_CMD(interp, "::CppVooPoint::set.name",  VooPoint, void, setName,  std::string);
    TCLCPPG_SETTER_CMD(interp, "::CppVooPoint::set.id",    VooPoint, void, setId,    int);
    TCLCPPG_SETTER_CMD(interp, "::CppVooPoint::set.active",VooPoint, void, setActive,bool);

    return Tcl_PkgProvideEx(interp, "VooPointCpp", "1.0.0", NULL);
}
\end{lstlisting}

\begin{table}[H]
\centering
\caption{Summary of migration steps for VOO C++}
\begin{tabular}{L{4.0cm}L{2.5cm}L{6.5cm}}
\toprule
\textbf{Step} & \textbf{Location} & \textbf{What it does} \\
\midrule
C++ class  & \texttt{point.cpp} & Plain C++ --- no Tcl headers required \\
\texttt{TypeName()} & \texttt{tcl::Type<>} specialization & Registers \texttt{VooPoint} in Tcl's type registry \\
\texttt{ToString()} & \texttt{tcl::Type<>} specialization & Enables \texttt{puts \$obj} and list/dict storage \\
\texttt{FromAny()}  & \texttt{tcl::Type<>} specialization & Parses a 5-element Tcl list back into a \texttt{VooPoint} \\
\texttt{VooPoint\_newDefault} & standalone command & Zero-argument constructor (\texttt{new()}) \\
\texttt{Point\_Init} & package init entry & Registers all Tcl commands; called by \texttt{load} \\
\bottomrule
\end{tabular}
\end{table}

The resulting Tcl API is identical to the pure-Tcl VOO declaration --- only the namespace
prefix (\texttt{CppVooPoint::} vs.\ \texttt{VooPoint::}) differs, so no caller code
needs to change when migrating from Tcl to C++.

\section{VOO Tcl Package}
\label{app:voopkg}

This appendix reproduces the complete source of \texttt{voo.tcl}~v1.0.0, the Tcl
implementation of the Vanilla Object Orientation framework described in the paper.
The listing is provided for reference and reproducibility.

\vspace{4pt}
\noindent\textbf{File:} \texttt{voo.tcl} \quad
\textbf{Version:} 1.0.0 \quad
\textbf{Lines:} 767

\vspace{6pt}

\lstset{numbers=left, numberstyle=\tiny\color{gray}, numbersep=8pt,
        stepnumber=5, firstnumber=1}

\begin{lstlisting}[language=Tcl, caption={\texttt{voo.tcl} --- complete source of the VOO Tcl package (v1.0.0)}]
# Vanilla Tcl Object Orientation (voo) package
namespace eval voo {
    # package version
    variable version 1.0.0

    variable handlerToObjectMap {}
    variable handlerCounter 0

    ##\brief Check if a namespace is a valid voo class
    # \param[in] namespaceName the namespace to check
    # \return 1 if valid voo class, 0 otherwise
    proc isVooClass {namespaceName} {
        if {![uplevel [list namespace exists $namespaceName]]} {
            return 0
        }
        return [expr {[uplevel [list namespace eval $namespaceName {
            info exists __defaultObj
        }]]}]
    }

    ##\brief Declare a new voo class namespace and process its class body
    # \param[in] args Arguments for class declaration: <className> <body> and optional -extends parent
    # \note Creates the class namespace, imports parent fields/methods when using -extends,
    #       and registers constructors and exports
    proc class {args} {
        set optDict {}
        set defaultArgs {}
        set numArgs [llength $args]
        for {set i 0} {$i < $numArgs} {incr i} {
            set arg [lindex $args $i]
            if {$arg eq "-extends"} {
                if {$i + 1 >= $numArgs} {
                    error "Constructor option '$arg' requires an argument"
                }
                dict set optDict $arg [lindex $args [incr i]]
            } elseif {$arg eq "-virtual" || $arg eq "-v"} {
                dict set optDict "-virtual" {}
            } else {
                lappend defaultArgs $arg
            }
        }
        lassign $defaultArgs className body
        set vooNs [namespace current]
        # create the namespace for the class
        uplevel [list namespace eval $className [subst -nocommands {
            namespace path [list $vooNs]
            variable __defaultObj {}
            variable __fields {}
            variable __tmp_isPublicEnabled 1
        }]]

        uplevel [list namespace eval $className {
            ##\brief Access default object for this class
            # \return Default class instance (list)
            # \note Used for inheritance and constructor defaults
            proc class.defaultObj {} {
                variable __defaultObj
                return $__defaultObj
            }
            
            ##\brief Get list of field names for this class
            # \return List of field names in declaration order
            # \note Useful for introspection and constructor -name new.args
            proc class.fields {} {
                variable __fields
                return $__fields
            }
        }]

        if {[dict exists $optDict -virtual] && [dict exists $optDict -extends]} {
            error "voo::class: cannot use -virtual with -extends; child classes inherit virtual automatically from a -virtual parent"
        }

        if {[dict exists $optDict -virtual]} {
            set normalizedClassName [uplevel [list namespace eval $className {namespace current}]]
            uplevel [list namespace eval $className [list variable __voo_is_virtual_class 1]]
            uplevel [list namespace eval $className [list variable __voo_class_namespace $normalizedClassName]]
            # Pre-populate __defaultObj with namespace tag at index 0 BEFORE field declarations
            # so that _getClassCurrNumFields returns 1 for the first field declared
            uplevel [list namespace eval $className [list set __defaultObj [list $normalizedClassName]]]
        }

        # variable __parentClassNamespace {}
        if {[dict exists $optDict -extends]} {
            set parentClassName [dict get $optDict -extends]

            if {![uplevel [list namespace exists $parentClassName]]} {
                error "Parent class '$parentClassName' does not exist."
            }

            # check if parent class exists
            if {![uplevel [list namespace eval $parentClassName {info exists __defaultObj}]]} {
                error "Parent class '$parentClassName' is not a valid voo class."
            }

            # normalize namespace name of parent class
            set parentClassName [uplevel [list namespace eval $parentClassName {
                namespace current
            }]]

            uplevel [list namespace eval $className [subst -nocommands {
                variable __parentClassNamespace $parentClassName
            }]]

            # import parent's default object values
            set parentDefaultObj [${parentClassName}::class.defaultObj]
            uplevel [list namespace eval $className [list set __defaultObj $parentDefaultObj]]

            # if parent is virtual, update namespace tag at index 0 to child's namespace
            set parentIsVirtual [uplevel [list namespace eval $parentClassName {info exists __voo_is_virtual_class}]]
            if {$parentIsVirtual} {
                set normalizedChildName [uplevel [list namespace eval $className {namespace current}]]
                uplevel [list namespace eval $className \
                    [list set __defaultObj [lreplace $parentDefaultObj 0 0 $normalizedChildName]]]
                uplevel [list namespace eval $className [list variable __voo_is_virtual_class 1]]
                uplevel [list namespace eval $className [list variable __voo_class_namespace $normalizedChildName]]
            }

            # import parent's field index variables by copying actual index values from parent
            set parentFields [${parentClassName}::class.fields]
            foreach field $parentFields {
                set fieldIdx [uplevel [list namespace eval $parentClassName [list set $field]]]
                uplevel [list namespace eval $className [list variable $field $fieldIdx]]
                uplevel [list namespace eval $className [list lappend __fields $field]]
            }

            # import parent's acessors in child class with namespace import
            uplevel [list namespace eval $className [subst -nocommands {
                namespace import ${parentClassName}::get.*
                namespace import ${parentClassName}::set.*
                namespace import ${parentClassName}::update.*
            }]]
        }

        uplevel [list namespace eval $className $body]

        uplevel [list namespace eval $className {
            if {[info commands new] eq ""} {
                constructor
            }
            if {[info commands new()] eq ""} {
                constructor -noargs [_buildConstructorNoArgsBody]
            }
            if {[info commands new.args] eq ""} {
                constructor -name new.args {args} [_buildConstructorArgsBody]
            }
        }]

        uplevel [list namespace eval $className {
            # export class methods
            namespace export *
        }]

        uplevel [list namespace eval $className {
            # clean temporary variable
            unset __tmp_isPublicEnabled
        }]
        return
    }

    ##\brief Return the default value for a given field type
    # \param[in] type the field type token (double,int,bool,...)
    # \return The default value appropriate for the type
    proc _getDefaultValueByType {type} {
        switch -- $type {
            double { return 0.0 }
            int    { return 0 }
            bool   { return 0 }
            default { return {} }
        }
    }

    ##\brief Get the current number of fields declared in the current class
    # \return Number of fields (integer)
    proc _getClassCurrNumFields {} {
        return [uplevel 2 {llength $__defaultObj}]
    }

    ##\brief Check whether public mode is enabled during class body parsing
    # \return 1 if public mode is enabled, 0 otherwise
    proc _getClassIsPublicEnabled {} {
        return [uplevel 2 {set __tmp_isPublicEnabled}]
    }

    ##\brief Declare getter/setter/updater accessors for a class field
    # \param[in] fieldName name of the field
    # \param[in] isPublic boolean whether accessors are public
    # \param[in] isStatic boolean whether field is static (class-level)
    proc _declareFieldAcessors {fieldName isPublic isStatic} {
        set prefix {}

        if {$isStatic} {
            append prefix class.
        }
        if {!$isPublic} {
            append prefix my.
        }

        set getterName "${prefix}get.$fieldName"
        set setterName "${prefix}set.$fieldName"
        set updaterName "${prefix}update.$fieldName"

        if {$isStatic} {
            uplevel 2 [list proc $getterName {} [subst -nocommands {
                variable $fieldName
                return $$fieldName
            }]]

            uplevel 2 [list proc $setterName {value} [subst -nocommands {
                variable $fieldName
                set $fieldName "\$value"
            }]]

            uplevel 2 [list proc $updaterName {tempVar body} [subst -nocommands {
                variable $fieldName
                upvar "\$tempVar" temp
                set temp $$fieldName
                # break link with class variable to avoid copy-on-write
                set $fieldName {}
                try {
                    uplevel \$body
                } finally {
                    set $fieldName "\$temp"
                }
            }]]
        } else {
            uplevel 2 [list getter $getterName $fieldName]
            uplevel 2 [list setter $setterName $fieldName]
            uplevel 2 [list updater $updaterName $fieldName]
        }
        return
    }
    
    ##\brief Validate a field name for illegal characters
    # \param[in] fieldName the field name to validate
    # \return Raises an error if invalid
    proc _validateFieldName {fieldName} {
        if {[string first "." $fieldName] != -1 || [string first "::" $fieldName] != -1} {
            error "Field name '$fieldName' cannot contain '.' or '::' substrings."
        }
    }

    ##\brief Ensure a field name does not already exist in the class
    # \param[in] fieldName the field name to check
    # \return Raises an error if the field already exists
    # \note Uses __fields for instance fields and fully-qualified namespace lookup for static
    #       fields to avoid false positives from global variables with the same name
    proc _validateFieldDoesNotExist {fieldName} {
        # Check instance fields tracked in __fields (class-scoped, no global bleed)
        if {$fieldName in [uplevel 2 {set __fields}]} {
            error "Field name '$fieldName' already exists in the class."
        }
        # Check static fields via fully-qualified namespace variable; info exists ::Ns::var
        # only matches that exact namespace variable, never a same-named global
        set classNs [uplevel 2 {namespace current}]
        if {[info exists ${classNs}::$fieldName]} {
            error "Field name '$fieldName' already exists in the class."
        }
    }

    ##\brief Validate a variable initial value according to its declared type
    # \param[in] type the declared type (double,int,bool,list,dict)
    # \param[in] value the value to validate
    # \return Raises an error if the value does not match the type
    proc _validateVarValueByType {type value} {
        switch -- $type {
            double {
                if {[string is double -strict $value] == 0} {
                    error "Value for t_double must be a double, got '$value'"
                }
            }
            int {
                if {[string is integer -strict $value] == 0} {
                    error "Value for t_int must be an integer, got '$value'"
                }
            }
            bool {
                if {[string is boolean -strict $value] == 0} {
                    error "Value for t_bool must be a boolean, got '$value'"
                }
            }
            list {
                if {[catch {llength $value}]} {
                    error "Value for t_list must be a list, got '$value'"
                }
            }
            dict {
                if {[catch {dict size $value}]} {
                    error "Value for t_dict must be a dict, got '$value'"
                }
            }
        }
    }

    ##\brief Declare a field variable inside the class body
    # \param[in] type the field type token (double,int,string,bool,list,dict,obj)
    # \param[in] argList arguments: ?-static? <name> ?<initialValue>?
    proc _var {type argList} {
        set defaultArgs {}
        set optDict {}
        set numArgs [llength $argList]
        for {set i 0} {$i < $numArgs} {incr i} {
            set arg [lindex $argList $i]
            if {$arg eq "-static"} {
                dict set optDict $arg {}
            } else {
                lappend defaultArgs $arg
            }
        }

        if {[llength $defaultArgs] == 0} {
            error "Variable definition requires: ?<option>? <name> ?<initialValue>?"
        }

        if {[llength $defaultArgs] == 2} {
            lassign $defaultArgs name initVal
        } else {
            lassign $defaultArgs name
            set initVal [_getDefaultValueByType $type]
        }

        _validateFieldName $name
        _validateFieldDoesNotExist $name
        _validateVarValueByType $type $initVal

        if {[dict exists $optDict -static]} {
            # static field
            uplevel [list variable $name $initVal]
        } else {
            set currNumFields [_getClassCurrNumFields]
            uplevel [list variable $name $currNumFields]
            uplevel [list lappend __defaultObj $initVal]
            uplevel [list lappend __fields $name]
        }

        set isPublicEnabled [_getClassIsPublicEnabled]
        _declareFieldAcessors $name $isPublicEnabled [dict exists $optDict -static]
        return
    }

    ##\brief Declare a double-typed field
    # \param[in] args same arguments accepted by _var (name and optional initial value)
    proc double_t {args} {
        uplevel [list _var "double" $args]
    }

    ##\brief Declare an integer-typed field
    # \param[in] args same arguments accepted by _var (name and optional initial value)
    proc int_t {args} {
        uplevel [list _var "int" $args]
    }

    ##\brief Declare a string-typed field
    # \param[in] args same arguments accepted by _var (name and optional initial value)
    proc string_t {args} {
        uplevel [list _var "string" $args]
    }

    ##\brief Declare a boolean-typed field
    # \param[in] args same arguments accepted by _var (name and optional initial value)
    proc bool_t {args} {
        uplevel [list _var "bool" $args]
    }
    
    ##\brief Declare a list-typed field
    # \param[in] args same arguments accepted by _var (name and optional initial value)
    proc list_t {args} {
        uplevel [list _var "list" $args]
    }

    ##\brief Declare a dict-typed field
    # \param[in] args same arguments accepted by _var (name and optional initial value)
    proc dict_t {args} {
        uplevel [list _var "dict" $args]
    }

    ##\brief Declare an object-typed field (nested vanilla object)
    # \param[in] args same arguments accepted by _var (name and optional initial value)
    proc obj_t {args} {
        uplevel [list _var "object" $args]
    }

    ##\brief Enable public mode for declarations inside the provided body
    # \param[in] body script to execute with public accessors enabled
    # \return Result of executing body
    proc public {body} {
        uplevel $body
    }

    ##\brief Execute the provided body with private mode enabled (temporarily disables public accessors)
    # \param[in] body script to execute with private accessors
    # \return Result of executing body
    proc private {body} {
        uplevel {variable __tmp_isPublicEnabled 0}
        try {
            uplevel $body
        } finally {
            uplevel {variable __tmp_isPublicEnabled 1}
        }
    }

    ##\brief Build the body for a no-argument constructor
    # \return A script chunk used as constructor body that returns the class default object
    proc _buildConstructorNoArgsBody {} {
        return {
            variable __defaultObj
            return $__defaultObj;
        }
    }
    
    ##\brief Build the body for a constructor that accepts named args (-field value pairs)
    # \return A script chunk used as constructor body that applies named arguments to the default object
    proc _buildConstructorArgsBody {} {
        return {
            variable __defaultObj
            set obj $__defaultObj
            if {[catch {dict size $args}]} {
                error "Constructor argument must be a list of '-<field> <value>' pairs"
            }
            dict for {key value} $args {
                if {[string index $key 0] ne "-"} {
                    error "Constructor argument keys must start with '-', got '$key'"
                }
                set field [string range $key 1 end]
                set setter set.$field
                if {[info commands $setter] ne ""} {
                    $setter obj $value
                } else {
                    set setter my.set.$field
                    if {[info commands $setter] ne ""} {
                        $setter obj $value
                    } else {
                        error "Unknown field option: $field"
                    }
                }
            }
            return $obj
        }
    }

    ##\brief Build constructor parameter list and body for positional constructors
    # \return A list of two elements: argument names list and a body script that returns them as a list
    # \note For virtual classes, the concrete class namespace is embedded as a literal string at
    #       class-definition time (not looked up at runtime), producing:
    #           return [list ::ClassName $f1 $f2 ...]
    #       This avoids all runtime proc calls (class.defaultObj, set.*) and variable lookups,
    #       making virtual object creation as cheap as non-virtual.
    proc _buildConstructorParams {} {
        set argList [uplevel 2 {set __fields}]
        set isVirtual [uplevel 2 {info exists __voo_is_virtual_class}]
        set spacedArgVarListStr {}
        foreach arg $argList {
            append spacedArgVarListStr "\$$arg "
        }
        if {$isVirtual} {
            # Read the normalized class namespace at definition time so subst embeds it
            # as a literal in the generated body - no runtime variable lookup required.
            set classNs [uplevel 2 {set __voo_class_namespace}]
            set spacedArgVarListStr "{$classNs} $spacedArgVarListStr"
            set body [subst -nocommands {
                return [list $spacedArgVarListStr]
            }]
        } else {
            set body [subst -nocommands {
                return [list $spacedArgVarListStr]
            }]
        }
        return [list $argList $body]
    }

    ##\brief Define a constructor for the current class
    # \param[in] args Constructor declaration options and body
    # \note Supports -name, -noargs and -typed variants
    proc constructor {args} {
        set defaultArgs {}
        set optDict {}
        set numArgs [llength $args]
        for {set i 0} {$i < $numArgs} {incr i} {
            set arg [lindex $args $i]
            if {$arg eq "-name" || $arg eq "-noargs" || $arg eq "-typed"} {
                if {$i + 1 >= $numArgs} {
                    error "Constructor option '$arg' requires an argument"
                }
                dict set optDict $arg [lindex $args [incr i]]
            } else {
                lappend defaultArgs $arg
            }
        }

        # check valid option combinations
        if {[dict exists $optDict -name]} {
            if {[dict exists $optDict -noargs] || [dict exists $optDict -typed]} {
                error "Constructor cannot have -name option with -noargs or -typed options"
            }
        }
        if {[dict exists $optDict -noargs] && [dict exists $optDict -typed]} {
            error "Constructor cannot have both -noargs and -typed options"
        }
        
        if {[dict exists $optDict -name]} {
            set constructorName [dict get $optDict -name]
        } elseif {[dict exists $optDict -noargs]} {
            set constructorName "new()"
        } elseif {[dict exists $optDict -typed]} {
            set constructorName "new([join [dict get $optDict -typed] ,])"
        } else {
            set constructorName "new"
        }

        if {[dict exists $optDict -noargs]} {
            if {[llength $defaultArgs] != 0} {
                error "Invalid constructor definition, expected '?...? ?<body>?' for -noargs"
            }
            set argList {}
            set body [dict get $optDict -noargs]
        } else {
            if {[llength $defaultArgs] == 0} {
                lassign [_buildConstructorParams] argList body
            } else {
                if {[llength $defaultArgs] != 2} {
                    error "Invalid constructor definition, expected '?...? ?<argList> <body>?'"
                }
                lassign $defaultArgs argList body
            }
        }

        uplevel [list proc $constructorName $argList $body]
        return
    }

    ##\brief Generate a getter procedure for a field
    # \param[in] methodName name of the generated getter (may include namespace prefix)
    # \param[in] fieldName name of the field to read
    proc getter {methodName fieldName} {
        # implementation of getter definition
        set fieldIdx [uplevel [list set $fieldName]]
        uplevel [subst -nocommands {
            ##\\brief Getter for $fieldName
            # \\param\[in\] this class instance
            # \\return $fieldName value
            proc $methodName {this} {
                return [lindex \$this $fieldIdx]
            }
        }]
        return
    }

    ##\brief Generate a setter procedure for a field
    # \param[in] methodName name of the generated setter (may include namespace prefix)
    # \param[in] fieldName name of the field to write
    proc setter {methodName fieldName} {
        # implementation of setter definition
        set fieldIdx [uplevel [list set $fieldName]]
        uplevel [subst -nocommands {
            ##\\brief Setter for $fieldName
            # \\param\[in\] thisVar name of variable containing class instance
            # \\param\[in\] value new value for $fieldName
            proc $methodName {thisVar value} {
                upvar \$thisVar this
                lset this $fieldIdx \$value
            }
        }]
        return
    }

    ##\brief Generate an updater procedure for a field (copy-on-write safe)
    # \param[in] methodName name of the generated updater (may include namespace prefix)
    # \param[in] fieldName name of the field to update by reference
    # \note The updater detaches the field to avoid unnecessary copying during updates
    proc updater {methodName fieldName} {
        # implementation of updater definition
        set fieldIdx [uplevel [list set $fieldName]]
        uplevel [subst -nocommands {
            ##\\brief Update $fieldName by reference
            # \\param\[in\] thisVar name of variable containing class instance
            # \\param\[out\] tempVar name of variable to hold $fieldName during update
            # \\param\[in\] body script to execute with $fieldName in tempVar
            # \\note Avoids copy-on-write by detaching field during update
            proc $methodName {thisVar tempVar body} {
                upvar \$thisVar this
                upvar \$tempVar temp

                set temp [lindex \$this $fieldIdx]
                # break link with object to avoid copy-on-write
                lset this $fieldIdx {}
                try {
                    uplevel \$body
                } finally {
                    lset this $fieldIdx \$temp
                }
            }
        }]
    }

    ##\brief Declare a method in the current class namespace
    # \param[in] args Method declaration arguments: name, argList, body and options (-static, -upvar, -update, -override)
    proc method {args} {
        set isPublicEnabled [_getClassIsPublicEnabled]
        set defaultArgs {}
        set optDict {}
        set numArgs [llength $args]
        for {set i 0} {$i < $numArgs} {incr i} {
            set arg [lindex $args $i]
            if {$arg eq "-static" || $arg eq "-upvar"} {
                dict set optDict $arg {}
            } elseif {$arg eq "-update"} {
                if {$i + 1 >= $numArgs} {
                    error "Method option '$arg' requires an argument"
                }
                dict set optDict $arg [lindex $args [incr i]]
            } elseif {$arg eq "-override"} {
                # Explicit override indicator
                dict set optDict $arg {}
            } elseif {$arg eq "-virtual"} {
                dict set optDict $arg {}
            } else {
                lappend defaultArgs $arg
            }
        }
        lassign $defaultArgs name argList body

        # check valid option combinations
        if {[dict exists $optDict -static]} {
            if {[dict exists $optDict -upvar] || [dict exists $optDict -update]} {
                error "Method cannot have both -static and -upvar or -update options"
            }
        }
        if {[dict exists $optDict -update]} {
            if {![dict exists $optDict -upvar]} {
                # automatically add -upvar if -update is specified
                dict set optDict -upvar {}
            }
        }

        set finalArgList {}
        set finalBody {}
        if {[dict exists $optDict -upvar]} {
            lappend finalArgList "thisVar" 
            append finalBody {
                upvar $thisVar this
            }
        } elseif {![dict exists $optDict -static]} {
            lappend finalArgList "this"
        }

        lappend finalArgList {*}$argList

        set className [uplevel {namespace current}]
        
        if {[dict exists $optDict -update]} {
            set updateFields [dict get $optDict -update]
            if {[llength $updateFields] == 0} {
                error "-update option requires at least one field name"
            }
            foreach field $updateFields {
                try {
                  set fieldIdx [uplevel [list set $field]]
                } trap {} {} {
                    error "Field '$field' specified in -update option does not exist in class '$className'"
                }
                append finalBody [subst -nocommands {
                    set $field [lindex \$this $fieldIdx]
                    lset this $fieldIdx {}
                }]
            }
            append finalBody "try \{"
        }
        append finalBody $body

        if {[dict exists $optDict -update]} {
            append finalBody "\} finally \{"
            foreach field $updateFields {
                set fieldIdx [uplevel [list set $field]]
                append finalBody [subst -nocommands {
                    lset this $fieldIdx \$$field
                }]
            }
            append finalBody "\}"
        }

        if {!$isPublicEnabled} {
            set name "my.$name"
        }

        if {[dict exists $optDict -override]} {
            set parentNs [uplevel {set __parentClassNamespace}]
            if {[info commands "${parentNs}::$name"] eq ""} {
                error "Method '$name' does not override any method in parent class '$parentNs'"
            }
            # If parent's method is virtual (has base.<name>), auto-promote this override
            # to a dispatcher so that deep inheritance dispatch works correctly
            if {[uplevel {info exists __voo_is_virtual_class}] && \
                    [info commands "${parentNs}::base.$name"] ne ""} {
                dict set optDict -virtual {}
            }
        }

        if {[dict exists $optDict -virtual]} {
            if {![uplevel {info exists __voo_is_virtual_class}]} {
                error "Method '$name' is declared -virtual but '[uplevel {namespace current}]' is not a virtual class"
            }
            if {[dict exists $optDict -upvar] || [dict exists $optDict -update] || [dict exists $optDict -static]} {
                error "Method '$name' cannot combine -virtual with -upvar, -update, or -static"
            }
            # Register base.<name> with the original body for direct parent calls from subclasses
            uplevel [list proc "base.$name" $finalArgList $finalBody]
            # Build dispatch body: route to concrete class implementation at runtime
            set dispatchBody "set __voo_cls \[lindex \$this 0\]\n"
            append dispatchBody "if \{\$__voo_cls ne \[namespace current\] && \[info commands \${__voo_cls}::$name\] ne {}\} \{\n"
            append dispatchBody "    return \[\${__voo_cls}::$name \$this"
            foreach arg $argList {
                append dispatchBody " \$$arg"
            }
            append dispatchBody "\]\n\}\n"
            append dispatchBody "return \[base.$name \$this"
            foreach arg $argList {
                append dispatchBody " \$$arg"
            }
            append dispatchBody "\]"
            set finalBody $dispatchBody
        }

        uplevel [list proc $name $finalArgList $finalBody]
        return
    }

    ##\brief Import one or more methods from parent class into the current (child) class namespace.
    # \param[in] methods List of method names (or a single method name) to import from parent.
    # \note Must be called inside a class declared with -extends. Methods are copied at class-definition time.
    proc importMethods {methods} {
        set parentNs [uplevel {set __parentClassNamespace}]

        # Validate caller context and get parent namespace stored by -extends handling
        if {$parentNs eq ""} {
            error "importMethods can only be used inside a class declared with -extends"
        }

        # Normalize to a list of method names
        if {[string length [string trim $methods]] == 0} {
            return
        }
        if {[catch {llength $methods}]} {
            set methodList [list $methods]
        } else {
            set methodList $methods
        }

        foreach methodName $methodList {
            set fullMethodName "${parentNs}::$methodName"
            # Validate parent method exists
            if {[info commands $fullMethodName] eq ""} {
                error "Method '$methodName' not found in parent class '$parentNs'"
            }

            # Define a copy in the child namespace so unqualified calls resolve to child
            set argList [info args $fullMethodName]
            set body [info body $fullMethodName]
            uplevel [list proc $methodName $argList $body]
        }
        return
    }

    namespace export *
}

# provide the package
package provide voo $voo::version
\end{lstlisting}

\lstset{numbers=none}

\end{document}